\documentstyle[12pt]{article}

      \newcommand{\beq}{\begin{equation}}
      \newcommand{\eeq}{\end{equation}}
      \newcommand{\beqa}{\begin{eqnarray}}
      \newcommand{\eeqa}{\end{eqnarray}}
      
      \newcommand{\nn}{\nonumber}
      \newcommand{\Tr}{{\rm Tr}}

      \newcommand{\be}{\beta}
      
      \newcommand{\de}{\delta}

      \newcommand{\bphi}{\mbox{\boldmath $\varphi$}}
\input epsf
\begin{document}
\begin{titlepage}
{\flushright UT-Komaba 96-1\\}
{\flushright INS-Rep.-1133\\}
{\flushright hep-th/9602086\\}

\makebox[1in]{      }
   \begin{Large}
       \begin{center}
         {\bf Convex effective potential of $O(N)$-symmetric } \\
         {\bf $\phi^4$ theory for large $N$}\\
        
       \end{center}
   \end{Large}
  \vfil
\begin{center}
         Hisamitsu Mukaida\footnote{\tt mukaida@ins.u-tokyo.ac.jp} \\
           \vspace{3mm}
           {\it Department of Physics, Saitama Medical College, Kawakado,
                Moroyama, Saitama, 350-04, Japan}\\
        \vspace{.5cm}
        and\\
        \vspace{.5cm} 
           Yujiro Shimada\footnote{\tt shimada@hep1.c.u-tokyo.ac.jp}\\
        \vspace{3mm}
        {\it Institute of Physics, University of Tokyo, 
              Komaba, Meguro-ku, Tokyo 153, Japan}\\
    \vfil
Abstract\\
  \end{center}
  We obtain effective potential of $O(N)$-symmetric $\phi^4$ theory for large $N$ 
starting with a finite lattice system and taking the thermodynamic limit with great care. 
In the thermodynamic limit, it is globally real-valued and convex in both the symmetric 
and the broken phases. 
In particular, it has a flat bottom in the broken phase.  
Taking the continuum limit, we discuss renormalization effects to the flat bottom and 
exhibit the effective potential of the continuum theory in three and four dimensions.
On the other hand the effective potential is nonconvex in a finite lattice system.  
Our numerical study shows that the barrier height of the effective potential flattens 
as a linear size of the system becomes large.  
It decreases obeying  power law and the exponent 
is about $-2$. 
The result is clearly understood from dominance of configurations 
with slowly-rotating field in one direction.   
\end{titlepage}
\section{Introduction}
Effective potential is widely employed in various contexts of physics 
for studying phase transition.
In the language of statistical physics, 
it is a free-energy density with some dynamical variable fixed. 
The fixed dynamical variable is no longer ``dynamical'',  and is called an order parameter.
A value of the order parameter in the thermodynamic limit specifies a phase of a model.
As shown in later section, the value of the order parameter is dynamically determined 
in a thermal equilibrium 
such that it gives a global minimum of the effective potential.  
Therefore we can recognize which phase is realized by looking at  
a form of the effective potential.

For example, let us consider  a classical ferromagnetic spin system.   
The order parameter in this case is a component of the spin variable 
averaged over all lattice sites.
According to the Landau-Ginzburg theory of phase transition, the effective potential 
is phenomenologically given as a polynomial of the order parameter: it  
has the unique minimum at the origin in the high-temperature (symmetric) phase. 
That is, a value of the order parameter is zero in this phase.   
As temperature is lowered, the effective potential  continuously changes and 
it eventually comes to have form of a double-well.  
Namely the order parameter does not vanish any more 
and we recognize that the system is in the low-temperature(broken) phase.

Although the Landau-Ginzburg theory gives a good qualitative picture of phase 
transition, the form of the phenomenological effective potential 
is  not globally correct because the effective potential of the spin system 
must be convex even in the low-temperature phase.  More precisely, as proved by 
O'Raifeartaigh et al., the effective potential of a  lattice scalar model 
with a nearest neighbor interaction plus  on-site potential 
 must be always convex even if the classical potential is not\cite{owy}. 
It means that a form like a double-well is ruled out. 
  
How does the effective potential look like  in the low-temperature phase? 
Numerical simulations for  $\phi^4$ theory on a lattice 
suggest that it has a flat bottom \cite{cm,owy}.  Similar effective potential 
is analytically obtained by Langer\cite{langer} in the spherical model 
with a help of the so-called  sticking argument\cite{bk,stanley}.
The flat region reflects dominance of configuration with domain structures\cite{hp,hs}, 
or dominance of spatially slowly varying configuration\cite{langer,rw} .    

A usual  way of computing the effective potential of $\phi^4$ theory 
is a  loop expansion from 
a uniform background, which is considered to be valid 
in a weak coupling region\cite{jackiw,coleman}. 
 It gives the classical potential in the leading  order.  
However, when the classical potential is not convex, the loop expansion 
develops an imaginary part 
due to  unstable modes of fluctuation\cite{fop}.  Systematic formulations  
that circumvent the unstable fluctuation have been proposed and one can obtain 
the (almost) convex effective potential by these methods\cite{fukuda,cf,rw,kms}.  
However, these results 
do not give us quantitative understanding because one cannot explicitly find  
any controllable expansion parameters.   Necessity of non-perturbative and quantitative 
approach naturally leads us to investigating the case where the number of fields, $N$, 
is sufficiently large and to carrying out the $1/N$ expansion. 

Investigation of phase structure of the large-$N$ $\phi^4$ theory using 
the effective potential has been extensively carried out for  more than two 
decades\cite{cjp,kobayashikugo,root,fukuda,bm,nunesch}.  
In order to determine a value 
of the order parameter by the effective potential, 
we need to know where a global minimum is. 
However,  most of the authors encountered 
trouble,   i.e., an imaginary part appears in the broken phase and they fails to obtain the 
effective potential globally.  In our knowledge, only exception is ref.\cite{fukuda}, 
where a non-uniform external field is used as a ``regularization".   
Although the effective potential seems to be globally obtained in that paper, 
the ansatz used to solve a saddle-point equation is incorrect. 
This point will be discussed more detail in sect.~5.

Thus, we attempt to find a global form of the effective potential of 
the $\phi^4$ model in the large-$N$ limit, which is the main subject of this paper.

We solve the saddle-point equation with giving a special attention to 
asymptotic property of the solution when the volume of the system 
tends to infinity.    
If one solves it after taking the thermodynamic limit,  
there are no real solutions for 
some range of order parameters in the broken phase.  This is the reason 
why the appearance of the imaginary part explained above.  
On the other hand, if one solves 
it in a finite  system and then takes the thermodynamic limit, 
one can obtain a globally real-valued effective potential. The idea of 
going back to a finite system  when one solves a saddle-point 
equation was used to reduce mathematical manipulation 
of the sticking argument\cite{lw,lautrup}.

The rest of the paper is organized as follows. 
In sect.~2, we define  $O(N)$-symmetric $\phi^4$ model on a lattice 
and give a formulation of the $1/N$ expansion of the effective potential 
in a finite lattice system. 
In sect.~3,  we  study asymptotic property of a solution of the saddle point 
equation derived in sect.~2 when the number of lattice sites of the 
system tends to infinity. Then we exhibit forms of the effective potential 
in the thermodynamic limit  for various dimensions.  
Also, we numerically solve the saddle-point equation 
in a finite system and quantitatively study 
how  effective potential in a finite system approaches 
to that in the thermodynamic limit. 
In sect.~4, we consider the continuum limit 
of the model and discuss the renormalization effects to the effective 
potential in three and 
four dimensions.   Section 5 is devoted to summary and discussion. 

Here we would like to comment on  definition of effective potential. 
The conventional way to define  the effective potential 
is to introduce an external source coupled to some dynamical variable.   
The expectation value of the dynamical variable, which is the order parameter in this case, 
and also the free-energy density,  are  a function of the external source.  We can regard the 
free-energy density  as a function of the order parameter by the Legendre 
transformation\cite{jackiw}.  The free-energy density constructed in this way 
is called the effective potential. 

Alternatively we can fix a dynamical variable by imposing  a constraint on a measure 
of the partition function so that the dynamical variable takes an given value(see eq.(\ref{cep})).  
The free-energy 
density derived from the partition function using this measure is also called the effective 
potential\cite{fk,hs} (or called the constraint effective potential\cite{owy}).

The relationship between the two is clarified in refs.\cite{hs,owy}: the former is the 
convex hull of the latter.  O'Raifeartaigh et al. also show that the effective potential 
of the lattice $\phi^4$ theory derived from the second definition is convex.  
Therefore the two definitions give the same effective potential in the case of 
the lattice $\phi^4$ theory. In this paper we adopt the second definition.

\section{The  $1/N$ expansion}
We study the $O(N)$-symmetric $\phi^4$ model on a $D$-dimensional 
hypercubic lattice.   The lattice spacing is set to be unity. 
There are $N$-scalar fields $\phi^a_{x}$, $a=1, \cdots, N$  
defined on each lattice site $x$.   We impose periodic boundary conditions: 
$\phi^a_x = \phi^a_{x + L e^\mu}$, $\mu = 1, \cdots, D$,  
where $e^\mu$ is the unit vector pointing to the $\mu$-direction and $L$ is an integer 
standing for the length of the side of the hypercubic lattice.  The model is defined by the action 
\beq
        S_0 = \sum_x  \{ \frac{1}{2} \phi^a_x (-\Delta + m_0^2)\phi^a_x  + 
                \frac{g_0}{8N}(\phi^a_x \phi^a_x)^2 \},
\label{s0} 
\eeq
where the summation over  $a$ is implicitly taken. The lattice Laplace operator 
$\Delta$ acts on  fields as 
\beq
        \Delta \phi^a_x =  \sum_{\mu=1}^D (\phi^a_{x + e_\mu}+\phi^a_{x - e_\mu} - 2 \phi^a_x).  
\eeq
Since we are  interested in the broken phase,   $m_0^2$ is set to be negative,  
while $g_0$ is taken to be positive because of the requirement that 
 $S_0$  should be bounded from below.   
  
We can replace  $S_0$ with  the following action by introducing an auxiliary field 
$\chi_x$ as follows\cite{cjp}:
\beq
        S = \sum_x  \{ \frac{1}{2} \phi^a_x (-\Delta + \chi_x)\phi^a_x - 
              \frac{N}{2 g_0} \chi_x^2 + \frac{N m_0^2}{g_0} \chi_x \}.
\eeq
In fact, the integration over $\chi_x$ in the partition function is easily performed 
and the resulting action equals $S_0$ up to a constant.  
Note that the sign of the $\chi_x^2$ term is negative:    
$\chi_x$ is integrated along the imaginary axis. 

 The effective potential $U$ of the finite system is defined by\cite{fk,cjp,owy} 
\beq
        \exp\{ -\Omega \ U(\Omega, \bphi^2) \} \equiv 
        \int \prod_{x,a} d \phi_x^a  \ \delta(\phi^a_c - \sqrt{N} \varphi^a) \ \prod_x d \chi_x 
        \exp \{ -S\}, 
\label{cep}
\eeq
where $\bphi^2 \equiv \varphi^a\varphi^a$ and $\Omega$ is the total number of 
the lattice sites.  We choose 
\beq
        \phi^a_c \equiv \frac{1}{\Omega} \sum_x \phi^a_x,  \qquad  a = 1, \cdots, N
\eeq
as order parameters. It is obvious that $U$ is a function of $\bphi^2$ due to the $O(N)$ symmetry.
 
The partition function $Z$  of the system is written in terms of $U(\Omega, \bphi^2)$ as 
\beq
 Z = N^{N/2} \int \prod_{a} d \varphi^a \exp \{ -\Omega \ U(\Omega, \bphi^2) \}.
\label{z}  
\eeq
In the limit of $\Omega \rightarrow \infty$, the integration on the right-hand-side 
of eq.(\ref{z}) is evaluated by a global minimum of  $U$. i.e., the global minimum of $U(\infty, \bphi^2)$ 
determines  a value of the order parameter realized in the thermodynamic limit. 
If there are multiple global minima, we need to choose 
one of them for looking at spontaneously symmetry breaking. A practical way of 
doing this is presented in sect.~3 (see also \cite{owy}). A physical meaning 
of this procedure is explained in the literature\cite{goldenfeld,wiedemann,kt}, 
for example.  

The classical potential 
\beq
        \frac{1}{2}m_0^2 \phi^a_x\phi^a_x  + \frac{g_0}{8 N} (\phi^a_x\phi^a_x)^2       
\eeq
takes the minimum on the $(N-1)$-dimensional sphere with  
the radius $\sqrt{- 2 N m_0^2/g_0}$ in $\mbox{\boldmath $\phi$}$-space(the target space). 
We expect that the values of $N \bphi^2$ minimizing $U$ lies in a range $N \bphi^2 \leq {\cal O}(N)$ 
as well as those of the classical potential, so that we shall search a minimum of $U$ in that  range:
$\bphi^2 \leq {\cal O}(N^0)$.

The main purpose of this section is to derive a formulation of the $1/N$ expansion of 
$U(\Omega, \bphi^2)$ following 
the spirit of refs.\cite{coleman,id}.  That is, we perform integration over $\phi_x$ and obtain a theory 
described by $\chi_x$.  The integration over $\chi_x$ is approximated by the saddle point method, which 
is applicable for large $N$.    

In order to carry out this program, 
we use an integral representation of the $\delta$-function in eq.(\ref{cep})
\beq
        \prod_a \delta(\phi^a_c - \sqrt{N} \varphi^a) = \int_{-i \infty}^{i \infty} \frac{d \eta}{2 \pi i \Omega} 
        \exp \{\eta^a (\sum_x \phi^a_x - \Omega \sqrt{N} \varphi^a)\}. 
\eeq
We can readily integrate over $\phi^a_x$ and $\eta^a$.  The result is, up to a constant, 
\beqa
        &&\exp\{ -\Omega \; U(\Omega, \bphi^2) \} = 
        \int \prod_x d \chi_x \exp \{ - N S_1 \}, \nn\\
        &&S_1[\chi] = \frac{1}{2}\Tr \ln (-\Delta +  \tilde \chi) +  \frac{1}{2} \ln f  +  
                     \sum_x (-\frac{1}{2 g_0} \chi_x^2 + \frac{m_0^2}{g_0} \chi_x ) + 
                     \frac{1}{2} \Omega^2 f^{-1} \bphi^2.
\label{s1}      
\eeqa
Here  we  defined the operator $\tilde \chi$  
\beq
         \tilde \chi |x \rangle = \chi_x |x \rangle, 
\eeq
and the functional $f$ 
\beq
                f \equiv \sum_{x,y} \langle x | (\frac{1}{-\Delta +  \tilde \chi}) | y \rangle.  
\eeq

The next step is to find the solution of the saddle-point equation 
\beq
        \frac{\de S_1[\chi]}{\de \chi_x} = 0.  
\label{originalsp}
\eeq
We assume translationally invariant solution $\chi^*$.   In this case, the above 
equation (\ref{originalsp}) is reduced to  
\beq
        \bphi^2 + \frac{2 m_0^2}{g_0} = 
        -\frac{1}{\Omega}\sum_{p \neq 0} \frac{1}{-\hat \Delta (p) + \chi^*} + 
        \frac{2}{g_0}\chi^*. 
\label{sp} 
\eeq

The solution $\chi^*$ is regarded as a function of $\bphi^2$ and $\Omega$.  
We get the leading-order effective potential, $U_0(\Omega, \bphi^2) $, by 
inserting $\chi^*$ into $S_1[\chi]$: 
\beqa
        U_0(\Omega, \bphi^2) &=& \frac{N}{\Omega} S_1[\chi^*] \nn\\
                                 &=&  \frac{N}{2 \Omega} \sum_{p \neq 0} 
                                         \ln (-\hat \Delta (p) + \chi^{*})- 
                                         \frac{N}{2 g_0} \chi^{* 2} + \frac{N m_0^2}{g_0}\chi^{*}  + 
                                         \frac{N}{2} \bphi^2  \chi^{*}.
\label{cep0} 
\eeqa
Further investigation to the leading order will be extensively carried out in later section. 

In the following we formulate calculation of  higher-order corrections. 
The $1/N$ expansion is the expansion around the saddle point.
Writing 
\beq
        \chi_x = \chi^* + i \frac{1}{\sqrt{N}} \xi_x, 
\eeq 
then we expand $S_1$ in terms of  $\xi_x$:  
\beq
        N S_1 [\chi] = N S_1[\chi^*] + \frac{N}{2} \sum_{x,y} 
                        \frac{\de^2 S_1}{\de \xi_x \de \xi_y}[\chi^*]  \xi_x \xi_y +
                         \sum_{n \geq 3}\frac{N}{n!} \; \frac{\de^n S_1}{\de \xi_{x_1} \cdots\de \xi_{x_n} }[\chi^*]  
                        \xi_{x_1} \cdots \xi_{x_n}.
\label{expansion}       
\eeq 
Note that a linear term in $\xi_x$ vanishes due to the saddle-point equation.

The explicit form of the quadratic term is computed using the following formulae:  
\beqa
        \Tr \ln(-\Delta + \tilde \chi) &=& \sum_{p} \ln(-\hat \Delta(p) + \chi^*) + 
        \frac{i}{\sqrt{N}\Omega}\sum_{p} \hat G_0(p) \sum_x \xi_x + \frac{1}{2 N} 
        \sum_{x,y} \xi_x \xi_y \Pi(x-y) + \cdots,  \nn\\
        f &=& \frac{\Omega}{\chi^*}\{1 - \frac{i}{\sqrt{N} \Omega \chi^* } \sum_{x} \xi_x - 
        \frac{1}{N \Omega \chi^*  } \sum_{x,y} \xi_x \xi_y G_0(x-y) \} + \cdots.
\eeqa
Here $\hat \Delta(p)$ is a Fourier component of the Laplace operator given by 
\beq
        \hat \Delta(p) = 2 (\sum_{\mu=1}^D \cos p_\mu - D), 
\eeq
and 
\beqa
        G_0(x) &\equiv& \frac{1}{\Omega}\sum_p e^{i p x} \hat G_0(p), \qquad 
        \hat G_0(p) \equiv \frac{1}{-\hat\Delta(p) + \chi^*}, \nn\\
        \Pi(x) &\equiv& \frac{1}{\Omega}\sum_p e^{i p x} \hat \Pi(p), \qquad 
        \hat \Pi(p) \equiv \frac{1}{\Omega} \sum_{q} \hat G_0(p-q) \hat G_0(q).
\eeqa
The result is  
\beqa
        N \frac{\de^2 S_1}{\de \xi_x \de \xi_y}[\chi^*]  &=& 
           \frac{1}{\Omega}\sum_{p} \hat D^{-1}(p)e^{i p (x-y)},\nn\\
        \hat D^{-1}(p) &=&  \frac{1}{2}(\hat \Pi(p) - \frac{1}{\Omega \chi^{* 2} }\de_{p \: 0}) 
        + \frac{1}{g_0} +  (\bphi^2 - \frac{1}{\Omega \chi^*}) (\hat G_0(p) - \frac{1}{\chi^*}\de_{p \; 0}).
\eeqa
We can read the propagator of $\xi_x$ from the quadratic term: 
\beq
        \langle \xi_x \xi_y \rangle = \frac{1}{\Omega} \sum_{p} e^{i p (x-y)} D (p). 
\eeq  

Integrating over $\xi_x$ with ignoring the cubic and higher-order vertices in (\ref{expansion}) gives 
the next-to-leading correction, which is of order $N^0$.  Higher-order corrections are calculated in 
a diagrammatic way.   
We see that the $p$-point vertex ($p \geq 3$)  is suppressed by the factor of $N^{-p/2+1}$.
Therefore a loop diagram that consists in $p_1$-, $\cdots$, $p_k$- point vertices contributes to 
the order of $N^{-(p_1 + \cdots + p_k)/2 + k }$.  The first nontrivial  
two-loop diagrams are generated by one 4-point vertex, or two 3-point vertices, which 
gives correction of order of $N^{-1}$.    

Finally we go back to the integration over $\phi^a_x$ that leads to eq.(\ref{s1}) and 
comment on an allowed range of $\chi^*$. 
The first term of $U_0(\Omega, \bphi^2)$ in eq.(\ref{cep0}) is derived from the integration 
over the oscillation modes of $\phi^a_x$.  The condition that the integration should converge  
is that 
\beq
        \delta + \mbox{Re} \ \chi^* > 0, 
\label{condition}
\eeq  
where $\delta$ is the smallest eigenvalue of $- \Delta$ in the oscillating modes:
\beq
        \delta \equiv 2 (1 - \cos(\frac{2 \pi}{L})). 
\label{delta}
\eeq

\section{Solution of the saddle point equation and a form of the effective potential}
\subsection{Effective potential in the thermodynamic limit}
In this subsection we shall study the leading-order effective potential of 
the lattice $\phi^4$ theory in the thermodynamic 
limit $\Omega \rightarrow \infty$.  In particular, we would like  to clarify  
the form of the effective potential in the broken phase.   
We denote 
\beq
        u_0 (\Omega, \bphi^2) \equiv \frac{1}{N} U_0 (\Omega, \bphi^2), \qquad
\label{u0}
\eeq
for convenience. Since $u_0(\Omega, \bphi^2)$ is an $O(N)$-symmetric function, 
we can put $\varphi^1 =t$, $\varphi^k =0 \: (2 \leq k \leq N)$ without loss of generality.
From eqs.(\ref{sp}) and (\ref{cep0}), we have 
\beq
        u_0'(\Omega, t^2) = \chi^* t, \\        
\label{first}
\eeq
\beq
u_0''(\Omega, t^2) = \chi^* + {\chi^{*}}' t,
\label{second}
\eeq
where ${}'$ means differentiation with respect to $t$.  
We wish to know asymptotic behavior of $\chi^*$ when $\Omega \rightarrow \infty$. 
Let 
\beq
        h(x) \equiv  
        -\frac{1}{\Omega}\sum_{p \neq 0} \frac{1}{-\hat \Delta (p) + x} + 
        \frac{2}{g_0} x. 
\label{h}       
\eeq
Namely, $h(\chi^*)$ is the right-hand-side of the saddle-point equation(\ref{sp}).  
The solution $\chi^*$ is 
read from the crossing point having the graph of $y = h(x)$ and a 
horizontal line representing $y = t^2 + 2 m_0^2/g_0$, as is shown in Fig.~1. 

Since $h(x)$ is 
strictly increasing and takes any real value 
for $x \in (-\de, \infty)$, 
there is a unique real solution satisfying (\ref{condition}) 
for arbitrary $t$ and $\Omega$.
\begin{figure}
  \begin{center}
    \begin{picture}(264,176)(0,-70)
     \put(0,0){\epsfxsize 264pt\epsfbox{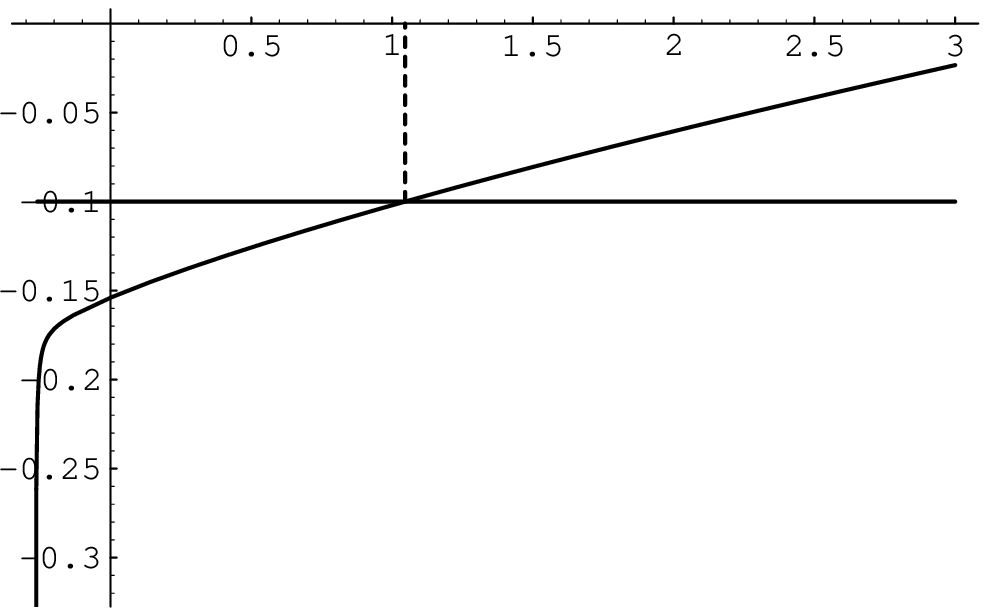}}
      \put(25,171){$y$}
      \put(105,171){$\chi^*$}
      \put(266,146){\small $y=h(x)$}
      \put(266,109){\small $y=t^2 + 2m_0^2/g_0$}
      \put(266,159){$x$}
      \put(0,-27){
                   \begin{minipage}[t]{264pt}
                     \begin{footnotesize}
                      Fig.~1. The graph of $h(x)$ in $D=4$, where $\Omega = 12^4$. 
                      The coupling constants are 
                      $g_0 = 80$, $m_0^2 = -15$. In this case, $-\delta \approx -0.26$.
                      There exists unique solution $\chi^*$ for all $t$.
                      \end{footnotesize}
                   \end{minipage}
                   }
    \end{picture}
  \end{center}
\end{figure} 
The sign of $\chi^*$ depends on whether the horizontal line lies above or 
below $h(0)$.  
If $\chi^*$ is  positive,  the summation over $p$ in 
$h(\chi^*)$ becomes the integration in the thermodynamic limit: 
\beq
        \lim_{\Omega \rightarrow \infty} h(\chi^*) = 
        - \int \frac{d^D p}{(2 \pi)^D}  \frac{1}{-\hat \Delta (p) + \chi^*} + 
        \frac{2}{g_0} \chi^*,
\label{inth} 
\eeq 
where the momentum $p_\mu$ ($1 \leq \mu\leq D$) runs from $-\pi$ to $\pi$. 
The saddle-point equation (\ref{sp}) in this case was essentially studied 
in ref.\cite{cjp}.  On the other hand, if $\chi^*$ is negative, we cannot employ 
eq.(\ref{inth}) any more and have to take the thermodynamic limit carefully.  
For this reason,  
we need to know asymptotic behavior of the graph of $h(x)$ in the neighborhood of $x=0$, 
which depends on the dimension of the system.    

For $D \leq 2$, $h(0)$ tends to $-\infty$ as $\Omega$ becomes large. 
Hence, for arbitrary $\bphi^2$, $\chi^*$ is  positive if $\Omega$ is sufficiently large.
Further,  it is shown that 
\beq
        {\chi^{*}}' \: t \geq  0 
\label{chi'}
\eeq 
because $h(x)$ is strictly increasing. 
Combining eqs.(\ref{first}), (\ref{second}) and (\ref{chi'})
we conclude that 
$u_0(\infty, \bphi^2)$ is strictly convex and has the unique minimum at the origin.
Therefore the order parameter vanishes in the thermodynamic limit. 
It is a typical feature  of the effective potential in the symmetric phase.  
The result corresponds to 
the fact that a continuous symmetry is not broken when $D \leq 2$ in a 
sense that a system has a non-vanishing order parameter.   

For $D \geq 3$,  $h(0)$ remains finite.  
If $x >0$ and  sufficiently small,  the integral in (\ref{inth}) 
is evaluated as follows\cite{id}:
\beq
        - \int \frac{d^D p}{(2 \pi)^D}  \frac{1}{-\hat \Delta (p) + x} \sim 
                \left\{
                        \begin{array}{ll}
                        -c_{3} + \frac{1}{4 \pi} \sqrt{x} & (D = 3)\\
                        -c_{4} -\frac{1}{16 \pi^2} x \ln x & (D = 4)\\
                        -c_{D} + K_{D} x & (D > 4)
                        \end{array} 
                \right.,
\label{smallchi}        
\eeq
where 
\beq
        c_{D} \equiv   
        \int \frac{d^D p}{(2 \pi)^D} \frac{1}{-\hat \Delta(p)}, \ \ (D \geq 3),   
\eeq
and $K_{D}$ is a non-universal positive constant. 

On the other hand, if $-\de < x < 0$, the graph of $h(x)$ drags a tail as seen in Fig.~1.  
Since $\de \rightarrow 0$ as $\Omega \rightarrow \infty$, 
the tail sticks along the $y$-axis, from $(0,h(0))$ towards $(0,-\infty)$ in  the 
thermodynamic limit.  This is the heart of the sticking argument\cite{bk,lw} used in the 
the spherical model\cite{stanley}.   We emphasize that we could not have found the tail 
and have lost a real solution for  $t^2 + 2 m_0^2/g_0 < h(0)$ if we took the thermodynamic 
limit {\em before} solving the saddle-point equation\cite{cjp,kobayashikugo,root,bm}. 

From the above consideration for the asymptotic behavior of $h(x)$, we conclude   
\beq
        \lim_{\Omega \rightarrow \infty} 
        \chi^* \  \left\{
                        \begin{array}{lll}
                                > & 0 & \ (t^2 >  \varphi_0^2)\\
                                = & 0 & \ (t^2 \leq \varphi_0^2) \\
                        \end{array}
                    \right.
\label{highdim}
\eeq
for $D \geq 3$.   Here we defined the constant $\varphi_0^2$ as 
\beq
        \varphi_0^2 \equiv -\frac{2 m_0^2}{g_0} -  c_D.
\label{phi0}  
\eeq
Note that $\varphi_0^2$ can be both positive and negative.   
The first term of the right-hand-side in eq.(\ref{phi0}) is related to  the radius 
of the sphere in the target space where the classical potential takes the minimum.  
The second term originates from statistical fluctuations.  If $\varphi_0^2 < 0$,  namely, 
if the statistical fluctuations dominate over the classical radius, 
the sphere ``disappears''  and the $O(N)$ symmetry will be restored.  

 In fact, if $\varphi_0^2 < 0 $, then $\chi^* > 0$ for all $t$.   
Namely the effective potential in this case has the unique minimum at the origin 
and the symmetry is not spontaneously broken  as in the lower dimensional case. 
 
In contrast,  if $\varphi_0^2 \geq 0 $, $\chi^*$ vanishes when $t^2 \leq \varphi_0^2$.    
This indicates that $u_0(\infty, \bphi^2)$ is  flat 
inside the $(N-1)$-dimensional sphere with the radius $\varphi_0$ in $\bphi$-space 
according to eq.(\ref{first}).  
This is an expected result in accord with numerical simulations\cite{cm,owy} or 
the Maxwell construction\cite{hp} when the symmetry is spontaneously broken. 
The reason why the flat region appears is that configurations with a domain 
structure contribute.\cite{hs,kms}.   

We can quantitatively study the form of $u_0(\infty, \bphi^2)$ outside of the flat region 
near the boundary, where we can assume $(\bphi^2- \varphi_0^2)$ is positive and sufficiently small.  
Since $0 < \chi^* << 1$ in this case and the formulae (\ref{smallchi}) are available.  
For $D =3$, we have 
\beq
        \chi^* = 16 \pi^2 (\bphi^2- \varphi_0^2)^2 + \cdots, 
\eeq
where $\cdots$ stands for terms higher than $ (\bphi^2 - \varphi_0^2)^2 $.  This provides, 
ignoring a constant term, 
\beqa
        u_0(\infty, \bphi^2)  &=& \frac{1}{2}\chi^* (\bphi^2 - \varphi_0^2) - 
                                  \frac{1}{12  \pi} \chi^{* 3/2} + \cdots, \nn\\
                              &=& \frac{8 \pi^2}{3} (\bphi^2 - \varphi_0^2)^3 + \cdots.   
\eeqa
Similarly, we have 
\beq
        u_0(\infty, \bphi^2) =   \left\{
                                        \begin{array}{ll}
                                                 4 \pi^2 (\bphi^2 - \varphi_0^2)^2 
                                                 (- \ln \{ 16\pi^2 (\bphi^2 - \varphi_0^2) \})^{-1}+ \cdots,
                                                & D=4  \\ 
                                                \frac{1}{4}(K_D + \frac{2}{g_0})^{-1} (\bphi^2 - \varphi_0^2)^2 + \cdots, 
                                                & D > 4
                                        \end{array}
                                        \right. .
\eeq

Using $u_0(\infty, \bphi^2)$, we can compute the one-point function in a usual  
manner\cite{fukuda,owy,id}.  
First we introduce a uniform external field $\sqrt{N} \mbox{\boldmath $J$}$ 
while $\Omega$ is kept finite.  That gives the following change in the partition function(\ref{z}):
\beq
        U_{0}(\Omega, \bphi^2) \rightarrow U_{0}(\Omega, \bphi^2) - N  J^a \varphi^a.
\eeq
The one-point function is computed as  
\beq
        \langle \frac{1}{\Omega} \sum_{x} \mbox{{\boldmath $\phi$}} \rangle =   
        \lim_{\mbox{{\boldmath $J$}} \rightarrow 0} \lim_{\Omega \rightarrow \infty}
        \frac{\sqrt{N}}{Z[ \mbox{{\boldmath $J$}}]} 
        \int \prod_a d \varphi^a \bphi \exp \{-\Omega \; N (u_0(\Omega, \bphi^2) -  J^a \varphi^a) \},
\label{onepoint} 
\eeq 
where $Z[ \mbox{{\boldmath $J$}}]$ is the partition function in the presence of the 
external field $\mbox{{\boldmath $J$}}$. 
Note that the limiting procedure does not commute. 
Since the external field explicitly breaks the $O(N)$ symmetry, $u_{0}(\infty,\bphi^2) - {\em J^a \varphi^a }$ 
takes the minimum at a unique point in $\bphi$-space, say, $\bphi^*$. 
The integration over $\varphi^a$ in (\ref{onepoint}) can be exactly evaluated at $\bphi^*$
in the thermodynamic limit.  
As $\mbox{\boldmath $J$} \rightarrow 0$,   $\bphi^*$ approaches to 
$\varphi_0$~\hspace{-.5mm}{\bf e}, where {\bf e} is the unit vector pointing to the same direction as 
the external field.  Thus we get 
\beq
        \langle \frac{1}{\Omega} \sum_{x}\mbox{{\boldmath $\phi$}} \rangle = \sqrt{N} \varphi_0 \; {\bf e}.  
\eeq
It shows that the symmetry is spontaneously broken.  

To summarize,  in the case of $D \geq 3$,  
the effective potential has a flat region and the order parameter 
takes the value of $\varphi_0$~\hspace{-.5mm}{\bf e} if and only if $\varphi_0^2 >0$.   
Namely, the symmetry is spontaneously broken if and only if 
$\varphi_0^2 >0$. 
The direction {\bf e} is determined by a uniform external field.  
\begin{center}
    \begin{picture}(211,141)(0,0)
      \put(0,0){\epsfxsize 211pt\epsfbox{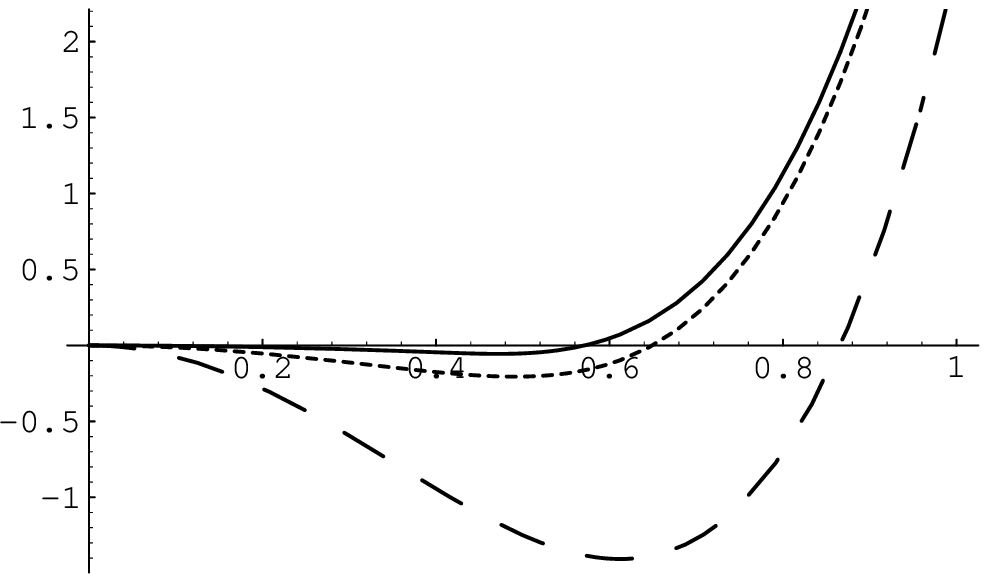}}
      \put(0,137){$u_0(\Omega, t^2)$}
      \put(215,53){$t$}
      \put(-20,-22){
        \begin{minipage}[t]{250pt}
         \begin{footnotesize}
        Fig.~2. The effective potential of a finite system  in $D=4$. 
        The classical potential(long-dashed line), $\Omega = 2^4$(dashed line) and 
        $\Omega= 8^4$ are presented. The coupling 
        constants are chosen as $g_0 = 80$, $m_0^2 = -15$, which are identical with 
        those of the Monte Carlo simulations in ref.\cite{owy} if we put $N=1$.
         \end{footnotesize}
        \end{minipage}
                 }
    \end{picture}
\end{center}
\vspace{4cm}
\subsection{Effective potential in a finite-size system}
In this subsection,   we show numerical results for  finite systems 
in order to see how the effective potential approaches to the form clarified 
in the previous subsection,  
as $\Omega \rightarrow \infty$.  We can numerically 
solve the saddle-point equation(\ref{sp}) for  finite systems and obtain Fig.~2.
A similar result is obtained in the context of renormalization-group 
analysis\cite{terao}.
We wish to  understand,   
from a quantitative point of view, how the barrier of the effective 
potential flattens as the system size becomes large.
To this end,  we define the barrier height $H$ as follows:
\beq
   H \equiv u_0(0, \Omega) -  \min u_0, 
\eeq
where we denote $\min u_0$ as a minimum value of the effective 
potential.  It is noted that an extreme value of the effective potential
satisfies the condition $\bphi=0$ or $\chi^*=0$ from eq.~(\ref{first}).
When the system is in the broken phase, $\chi^*$ is negative
at $\bphi=0$ for all $\Omega$, so that $u_0$ has a local maximum at $\bphi=0$.
On the other hand, 
since $h(x)$ of eq.~(\ref{h}) is strictly increasing as discussed above,
there exists a solution $\chi^* = 0$ for some $|\bphi |^2 > 0$, 
where $u_0$ has a local minimum.
Thus, from eqs.(\ref{cep0}) and (\ref{u0}),  
the barrier height $H$ is  explicitly written as  
\beq
    H   = \frac{1}{2 \Omega} \sum_{p \neq 0} \{
                                         \ln (-\hat \Delta (p) + \chi^{*}(0))-
 \ln (-\hat \Delta (p))
                   \} - 
                                         \frac{1}{2 g_0} \chi^{* 2}(0) + 
      \frac{ m_0^2}{g_0}\chi^{*} (0),  
\label{H} 
\eeq
where 
\beq
   \chi^*(0) \equiv \chi^*|_{\bphi = 0}.
\eeq  

 Let us first consider how $H$ depends on the coupling 
constants in a finite-size system.  Since $-\de < \chi^*(0) < 0$ in the broken
 phase, 
$\chi^*(0)$ goes to zero as $\Omega \rightarrow \infty$.  Therefore we can 
neglect the $\chi^{* 2}(0)$-term in eq.~(\ref{H}) and also, due to the same 
reason, 
the $\chi^*$-term in eq.~(\ref{sp}).  That means that $H$ depends only on 
the combination $m_0^2/g_0$.  
Neglecting these terms can be effectively 
carried out by taking the limit $g_0 \rightarrow \infty$ with 
$\beta \equiv - m_0^2/g_0$ kept
 finite. 
In this limit the $O(N) \ \phi^4$ model becomes the $O(N)$ nonlinear 
$\sigma$-model, where a length of a field variable is fixed to
$\sqrt{2 N \beta}$.

We numerically compute the barrier height varying coupling constants 
with $\beta$ kept fixed and 
plot the ratio of these barrier heights
to that of the $O(N)$ nonlinear $\sigma$-model 
as a function of $L$ in $D=3$(Fig.~3) and $D=4$(Fig.~4).
\vspace{2cm}
\begin{center}
    \begin{picture}(211,157)(0,0)
      \put(0,0){\epsfxsize 211pt\epsfbox{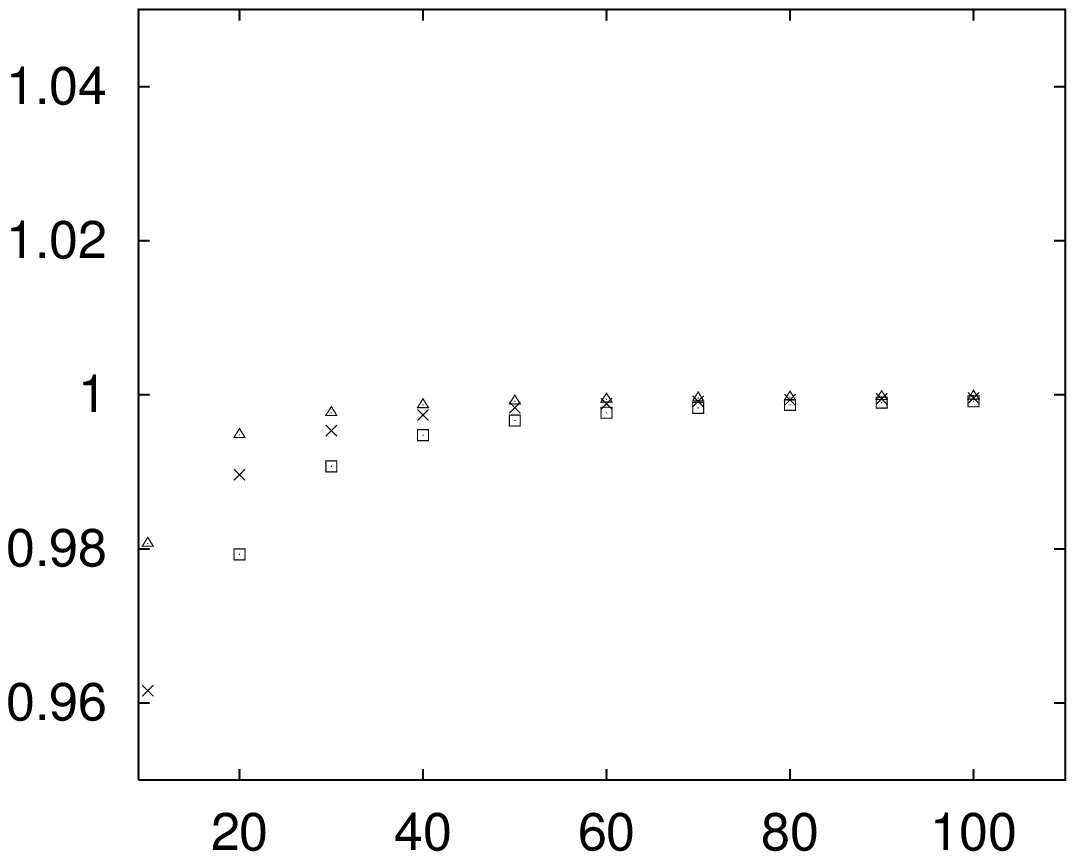}}
      \put(-44,136){$H/H_0$}
      \put(215,5){$L$}
      \put(-20,-22){
                   \begin{minipage}[t]{250pt}
                     \begin{footnotesize}
                      Fig.~3. The ratio of the barrier height
 $H/H_0$ as a function of $L$ 
from $L=10$ to $100$ in $D=3$.
 The coupling constants are 
 $(g_0,m_0^2) =(40,\,-7.5)$(boxes), 
$(g_0,m_0^2) =(80,\,-15)$(crosses),
and $(160,\,-30)$(triangles), where $\beta$ is kept at 0.1875. 
$H_0$ is the barrier height of the $O(N)$ nonlinear $\sigma$-model. 
                      \end{footnotesize}
                   \end{minipage}
                   }
    \end{picture}
\end{center}
\newpage
\begin{center}
    \begin{picture}(211,157)(0,0)
      \put(0,0){\epsfxsize 211pt\epsfbox{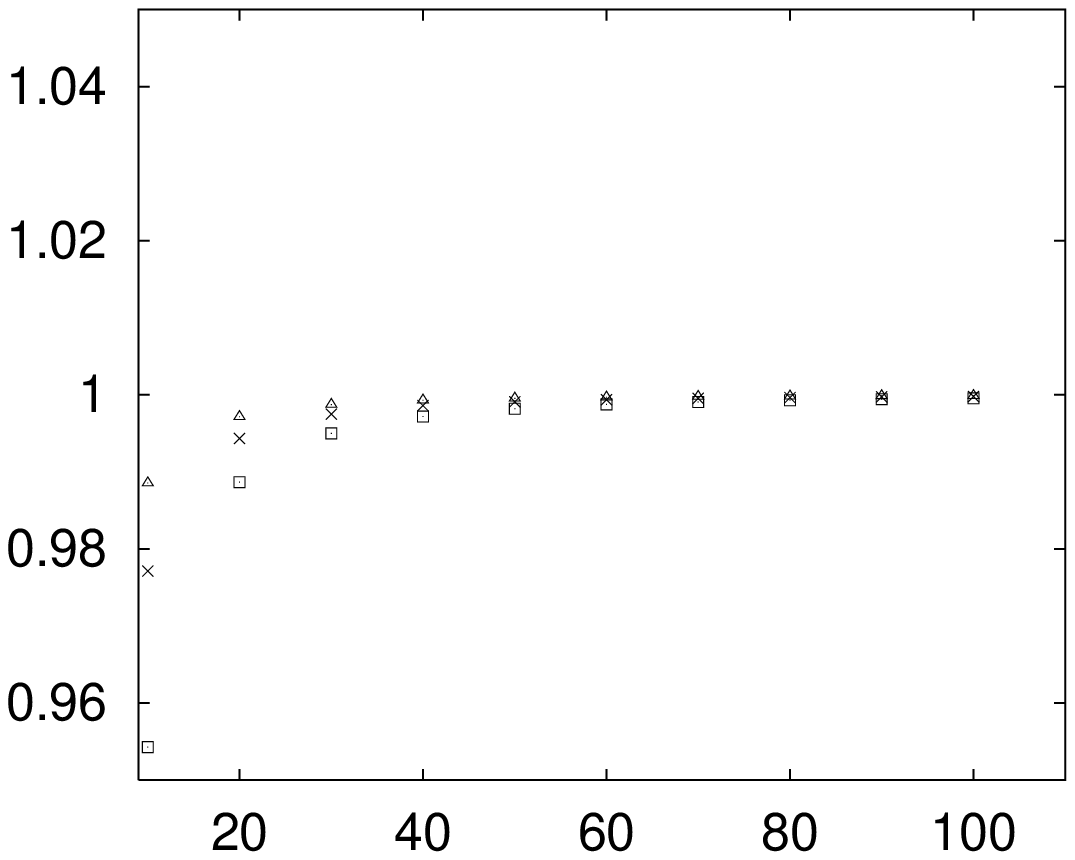}}
      \put(-44,136){$H/H_0$}
      \put(215,5){$L$}
      \put(-20,-22){
                   \begin{minipage}[t]{250pt}
                     \begin{footnotesize}
                      Fig.~4. The ratio of the barrier height
 $H/H_0$ as a function of $L$ 
from $L=10$ to $100$ in $D=4$.
 The coupling constants are 
 the same as Fig.~3.
                     \end{footnotesize}
                   \end{minipage}
                   }
    \end{picture}
\end{center}
\vspace{4cm}
The result  shows that they have the same asymptotic behavior 
as that of the $O(N)$ nonlinear $\sigma$-model for sufficiently 
large $L$, as we expected.  
Hence we conclude that, when the system is in the broken phase, 
the length of the field at each lattice is 
$\sqrt{2 N \beta}$ in dominant 
configurations for sufficiently 
large $L$.
 
In order to clarify dominant configuration at $\bphi = 0$ more definitely,  
we treat $H$ as a function of $L$ and study its asymptotic behavior.  
When the parameters  $\beta$ and $L$ 
are sufficiently large,  contribution from energy to the effective potential 
will dominate over that from entropy in the broken phase.  If the energy 
density (i.e., the value of the action per unit volume) 
 of most dominant configurations at $\bphi = 0$ behaves as 
\beq
        {\rm min}~u_0 + {\rm const.} \times L^{-\gamma},  
\label{gamma}  
\eeq
then $H$ will indicate  asymptotic behavior as 
${\rm const.} \times L^{-\gamma}$.  

The barrier heights as a function of $L$ in $D=3$ and $D=4$ are shown in Fig.~5.  
We find that for $L \geq 10$ each $H$ is fitted very well by the power function 
eq.(\ref{gamma}).  The fitting exponent is summarized in Table~1.  The result 
is that the all exponents are very close to 2.  
\newpage
\begin{center}
    \begin{picture}(316,235)(0,-50)
      \put(0,0){\epsfxsize 316pt\epsfbox{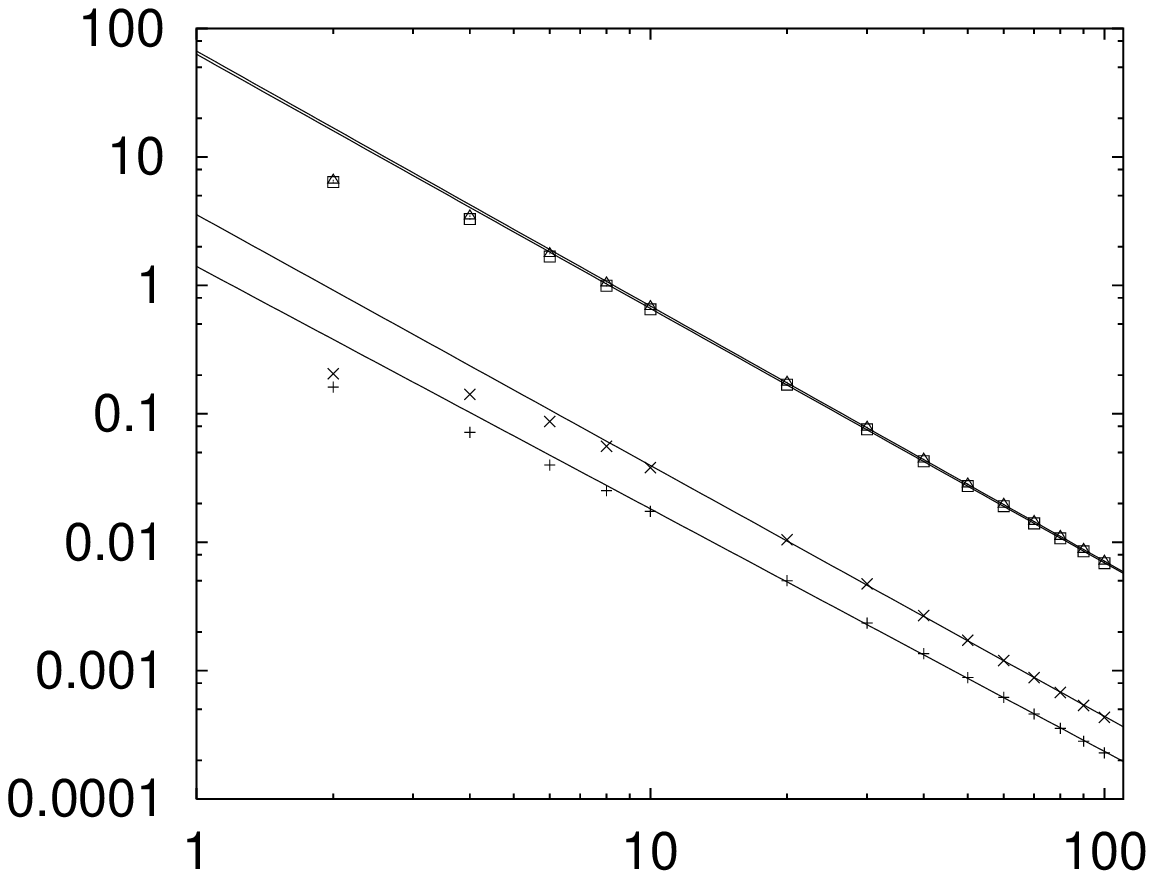}}
      \put(-10,204){$H$}
      \put(322,8){$L$}
      \put(10,-33){
                   \begin{minipage}[t]{316pt}
                     \begin{footnotesize}
                      Fig.~5. The barrier height
 $H=u_0(0,\Omega)-\min u_0$ as
                      a function of $L$ 
from $L=2$ to $100$ with
$(g_0,m_0^2) = (80,-15)$ in $D=3$(daggers),
$(g_0,m_0^2) = (80,-150)$ in $D=3$(boxes),
$(g_0,m_0^2) = (80,-15)$ in $D=4$(crosses),
$(g_0,m_0^2) = (80,-150)$ in $D=4$(triangles). 
Each line is a fitting power function.
                      \end{footnotesize}
                   \end{minipage}
                   }
    \end{picture}
\end{center}
\vspace*{3cm}
\begin{table}[h]
 \begin{center}
   \begin{tabular}{|l|l|c|}  \hline 
       $D$ & $(g_0, m_0^2)$ & $\gamma$ \\ 
       \hline
       3       & $(80, -15)$  & 1.888  \\ 
       3       & $(80, -150)$ & 1.980  \\ 
       4       & $(80, -15)$ & 1.953  \\ 
       4       & $(80, -150)$ & 1.986  \\
       \hline
   \end{tabular}
   \caption{Fitting exponent $\gamma$ in the case of 
$(g_0,m_0^2) = (80,-15), (80,-150)$ in $D=3$ and 4.}
 \end{center}
\end{table}
\newpage
In order to understand the above asymptotic property, 
let us find a configuration satisfying the following conditions: 
\begin{enumerate}
        \item{$\bphi = 0$, }
        \item{the length of the field at each lattice is $\sqrt{2 N \beta}$, }
        \item{its energy density  behaves as ${\rm const.} \times L^{-2}$.}
\end{enumerate}

Let us observe a configuration where $\phi^a_x$ with length $\sqrt{2 N \beta}$ slowly rotates 
for one direction and is uniform for all the other directions in the lattice.  For example\footnote{Similar configuration was considered by 
Ringwald and Wetterich\cite{rw}. }, 
\beq
        \phi^1_x = \sqrt{2 N \be } \cos \frac{2 \pi x^1}{L}, \ \ \ 
        \phi^2_x = \sqrt{2 N \be } \sin \frac{2 \pi x^1}{L}, \ \ \ 
        \phi^i_x = 0 \ \ \ (3 \leq i  \leq N).
\label{rotate}  
\eeq
We find from eq.~(\ref{s0}) that  
the energy density of this configuration 
measured from that of the uniform configuration $\phi^1_x = \sqrt{2 N \be }$ is 
$N \beta \delta$, which asymptotically behaves as  $L^{-2}$.  
Hence  
such configurations satisfy the above conditions.

 It is noted that  energy density of a configuration 
composed of some macroscopic uniform domains
can be roughly estimated to be proportional to $L^{-1}$.

Therefore our numerical study strongly suggests that the barrier height 
of the effective potential 
in a finite-size system flattens as the system size becomes large 
because the slowly rotating configuration described above becomes 
dominant as the size becomes large. 

%
%
\section{The continuum limit and renormalization effects}
So far, we have treated the effective potential of the lattice model.  
In this section, we take the continuum limit and study 
the renormalization effects. 
The global form of the effective potential of the continuum theory 
in dimension lower than three is obtained in  
refs.\cite{coleman,root}.  Here we restrict the dimension of the 
system to three or four.  
Our main interest is to see a fate of the flat region.  

In order to take the continuum limit, we introduce the lattice 
spacing parameter $a$,  which measures dimensions of physical 
quantities.  We define  
\beqa
        V &\equiv& a^D \Omega, \nn\\
        \varphi_{\rm R}^b &\equiv& a^{-\frac{D-2}{2}} \varphi^b, \nn\\
        \phi_{\rm R}^b &\equiv& a^{-\frac{D-2}{2}} \phi^b, \nn\\
        \chi^*_{\rm R} &\equiv& a^{-2} \chi^*,
\label{ren} 
\eeqa
where we assign the canonical dimension to the scalar fields. 
The effective potential of the lattice theory with the finite volume $V$ and with 
the lattice parameter $a$,  
$u_{\rm R}(a, V,\bphi_{\rm R}^2)$,  is defined by 
\beq
        u_{\rm R}(a, V, \bphi_{\rm R}^2) \equiv a^{-D} u_0(\Omega, \bphi^2).
\eeq
It is explicitly written in terms of dimensionful quantities defined in eq.(\ref{ren}).  Using 
eqs.(\ref{cep0}) and (\ref{u0}), we get  
\beq
           u_{\rm R}(a, V, \bphi_{\rm R}^2) 
        = \frac{1}{2} \bphi_{\rm R}^2  \chi^{*}_{\rm R}  - 
           \frac{1}{2 g_0(a) a^{D-4}} \chi^{* 2}_{\rm R} + \frac{m_0^2(a) a^{-2}}{g_0(a)a^{D-4}}\chi^{*}_{\rm R} +
              \frac{1}{2 V} \sum_{\bar p \neq 0} \ln (-\hat \Delta (\bar p \; \! a) a^{-2} + \chi^{*}_{\rm R}).
\label{cepren} 
\eeq
Here 
$ \chi^{*}_{\rm R} $ is 
the solution of the saddle-point equation of the theory with 
the lattice spacing $a$, which is derived from eq.(\ref{sp}):  
\beq
        \bphi_{\rm R}^2 + \frac{2 m_0^2(a) a^{-2}}{g_0(a) a^{D-4}} = 
        -\frac{1}{V}\sum_{\bar p \neq 0} 
           \frac{1}{-\hat \Delta(\bar p \; \! a) a^{-2} + \chi^*_{\rm R} } 
        + \frac{2}{g_0(a) a^{D-4}} \chi_{\rm R}^{*},
\label{spren} 
\eeq
where the momentum $\bar p_{\mu}$ is quantized by $2 \pi V^{-1/D}$ and runs from 
$-\pi/a$ to $\pi/a$.    
We adopt a renormalization prescription to keep $u_{\rm R}(a,V,\bphi^2_{\rm R})$ finite 
in the continuum limit $a \rightarrow 0$.  This determines lattice-spacing 
dependence of the bare parameters, $g_0(a)$ and $m^2_0(a)$.

We shall first consider the case of  $D=3$. 
The bare  parameters  are chosen in such a way 
that the following equations hold\cite{cjp}: 
\beqa
        g_{\rm R} &=& g_0(a) a^{-1}, \nn\\
        \frac{2 m_{\rm R}^2}{g_{\rm R}}
        &=& \frac{2 m^2_0(a) a^{-2}}{g_{0}(a) a^{-1}} + 
        \frac{1}{V}\sum_{\bar p \neq 0} 
         \frac{1}{-\hat \Delta(\bar p \; \! a) a^{-2}},
\label{cren}
\eeqa
where we introduced the renormalized mass and the coupling constant,  
$m^2_{\rm R}$ and $g_{\rm R}$. 
They are directly related to physical observables of the continuum theory.  
The continuum limit $a \rightarrow 0$ is taken keeping $\varphi_{\rm R}^b$, 
$m_{\rm R}^2$, $g_{\rm R}$, and $V$ finite.  
In this limit, we can make the following substitution in (\ref{cepren}) and (\ref{spren})  
\beq
         -\hat \Delta(\bar p \; \! a) a^{-2} + \chi^*_{\rm R} \rightarrow 
         {\bar p}^2 + \chi^*_{\rm R}, 
\eeq
where $ {\bar p}^2\equiv \sum_{\mu} \bar p_{\mu} \bar p_{\mu}$.  
The continuum effective potential is written 
in  terms of the renormalized coupling constants 
\beq
 u_{\rm R}(0, V, \bphi_{\rm R}^2) 
        = \frac{1}{2} \bphi_{\rm R}^2  \chi^{*}_{\rm R}  - 
           \frac{1}{2 g_{\rm R}} \chi^{* 2}_{\rm R} + \frac{m_{\rm R}^2}{g_{\rm R}}\chi^{*}_{\rm R} +
              \frac{1}{2 V} \sum_{\bar p \neq 0} \{ \ln (\bar p^2 + \chi^{*}_{\rm R} ) -  
           \frac{1}{{\bar p}^2 } \chi^{*}_{\rm R} \}, 
\label{cep3} 
\eeq
where $\chi^*_{\rm R}$ satisfies 
\beq
        \bphi_{\rm R}^2 + \frac{2  m^2_{\rm R}}{g_{\rm R}} = 
        -\frac{1}{V} \sum_{\bar p \neq 0} ( \frac{1}{\bar p^2+ \chi^*_{\rm R}}
        - \frac{1}{\bar p^2} ) + \frac{2}{g_{\rm R}} \chi^*_{\rm R}.
\label{sp3}
\eeq

Our aim is to clarify the global form of
 the effective potential in the infinite-volume limit.  
Behavior of the solution in the limit $V \rightarrow \infty$ is the same as that 
in the lattice model.  Namely,  
if $m_{\rm R}^2 >  0$, 
\beq
        \lim_{V \rightarrow \infty } \chi^*_{\rm R} > 0, 
\eeq
and if $m_{\rm R}^2 <  0$, 
\beq
        \lim_{V \rightarrow \infty} 
        \chi^*_{\rm R} \  \left\{
                        \begin{array}{lll}
                                > & 0 & \ (\bphi^2_{\rm R}  >  - 2  m_{\rm R}^2/g_{\rm R})\\
                                = & 0 & \ (\bphi^2_{\rm R}  <  - 2  m_{\rm R}^2/g_{\rm R}) \\
                        \end{array}
                    \right..
\eeq
We conclude from this behavior that $u_{\rm R}(0,\infty, \bphi^2_{\rm R})$ is flat inside the sphere 
with the radius $\sqrt{- 2 m_{\rm R}^2/g_{\rm R}}$ when the symmetry is broken
($m_{\rm R}^2 < 0$),  as is the case of the lattice model. 
When $\chi^*_R >0$, the summation over $\bar p$ in (\ref{sp3}) is replaced by 
the integration and the result is 
\beq
        \bphi_{\rm R}^2 + \frac{2  m^2_{\rm R}}{g_{\rm R}} = 
        \frac{1}{4 \pi} \sqrt{\chi^*_{\rm R}} 
        + \frac{2}{g_{\rm R}} \chi^*_{\rm R}.
\eeq
We can explicitly solve the equation.  Inserting the solution into (\ref{cep3}), 
we obtain $u_{\rm R} (0,\infty,\bphi^2_{\rm R})$ outside the flat region. 
We depict the result in Fig.~6.  
\begin{figure}[t]
  \begin{center}
    \begin{picture}(264,196)(0,-68)
      \put(0,0){\epsfxsize 264pt\epsfbox{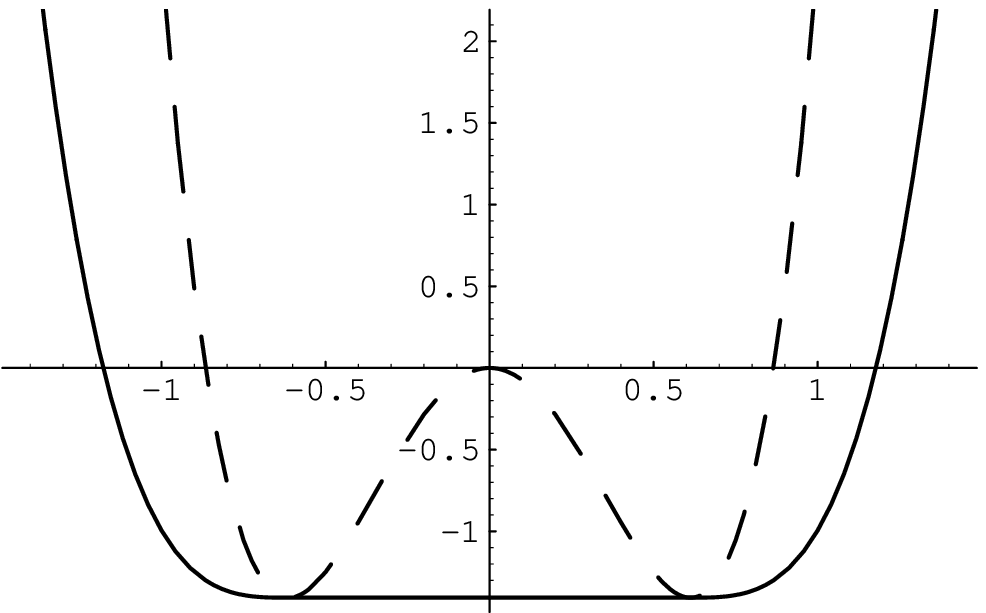}}
      \put(98,171){$u_{\rm R}(\infty, t^2)$}
      \put(269,66){$t$}
      \put(0,-60){
        \begin{minipage}[h]{264pt}
         \begin{footnotesize}
        Fig.~6. The effective potential of the continuum theory in $D=3$ 
        and the classical potential.
        The renormalized coupling constants are chosen as $m^2_{\rm R} = -15$
        and $g_{\rm R} = 80$.
         \end{footnotesize} 
        \end{minipage}
         }
    \end{picture}
  \end{center}
\end{figure}

It may possible to take 
other limiting procedures for obtaining $u_{\rm R} (0, \infty, \bphi^2_{\rm R})$.  
For example,  one may first take the limit $\Omega \rightarrow \infty$ 
with fixed  $a>0$ and next take the limit $a \rightarrow 0$.  
It corresponds to taking the infinite-volume limit {\em before} taking 
the continuum limit.   
We can obtain the same result in this limiting procedure if we solve 
the saddle-point equation before taking the limit $\Omega \rightarrow \infty$. 

Let us turn to the case of $D=4$.  We introduce the renormalized coupling 
constants by the following equations\cite{cjp}
\beqa
        \frac{2}{g_{\rm R}} &=& \frac{2}{g_0(a)} + \frac{1}{V} \sum_{\bar p \neq 0}
                                      \frac{1}{(-\hat \Delta(\bar p \; \! a) a^{-2} + M^2)
                                            ( -\hat \Delta(\bar p \; \! a) a^{-2}) }, 
                             \nn\\
        \frac{2 m_{\rm R}^2}{g_{\rm R}} &=& \frac{2m_0^2(a) a^{-2}}{g_0(a)} + 
                                                \frac{1}{V} \sum_{\bar p \neq 0} 
                                                \frac{1}{-\hat \Delta(\bar p \; \! a) a^{-2}},  
\label{re4}
\eeqa
where $M^2$ is an arbitrary positive constant.  
The saddle-point equation can be written in the finite form as 
\beq
        \bphi^2_{\rm R} + \frac{2m_{\rm R}^2}{g_{\rm R}} = \frac{\chi^*_{\rm R}}{V} \sum_{\bar p \neq 0}
                                                        \frac{1}{\bar p^2} 
                                                      ( \frac{1}{\bar p^2+ \chi^*_{\rm R}} 
                                                        - \frac{1}{\bar p^2 + M^2} ) + 
                                                     \frac{2}{g_{\rm R}} \chi^*_{\rm R}.        
\eeq

Since the summation over $\bar p$ in the first equation in eq.(\ref{re4})
diverges as $a \rightarrow 0$, $g_{\rm R}$ must be chosen to zero 
as long as we start 
from a positive $g_0(a)$\cite{cjp,bm}. This result corresponds to the conjecture that the $\phi^4_4$ theory
becomes trivial in the continuum limit.   The saddle-point equation becomes 
\beq
        m^2_{\rm R} = \chi^*_{\rm R}. 
\eeq
Namely, the theory reduces to the Gaussian model.  
The effective potential in the continuum limit is 
\beq
        u_{\rm R}(0,V, \bphi^2_{\rm R}) = \frac{1}{2} m_{\rm R}^2 \bphi^2_{\rm R},  
\eeq
up to an irrelevant additive constant. 
We must choose 
$m_{\rm R}^2$ to be non-negative for well-defined continuum filed theory.
%
%
\section{Summary and discussion}
In this paper, we obtained the effective potential of the lattice 
$\phi^4$ theory in the large-$N$ limit.  In the thermodynamic limit, we 
 found out that it is globally 
real-valued and convex in both the symmetric and the broken phases 
even if classical potential is not convex.  
In particular, the effective potential has a flat region in the broken phase, 
which  is consistent with numerical simulations and with the Maxwell construction. 

The effective potential in a finite system has a potential barrier at the origin although 
it is much lower than that of the classical potential.  
The height of the barrier is 
related to energy of defects of non-uniform configurations,  and asymptotically flattens 
as $L^{-2}$, where $L$ is the linear size of 
the system.  The result concerning with a finite lattice system 
can be applied to checking the validity of an $1/N$ expansion\cite{flyvbjerg}.
The effective potential with sufficiently flat maximum as shown in Fig.~2 
may shed light on global structure of inflationary universe\cite{alindedlinde}.

We also studied the continuum limit of this model and discussed 
effects of the renormalization to the flat region in the case of 
$D=3$ and $D=4$.  In the case of $D=3$, the effective potential 
in the continuum limit has the flat region as in the case of the 
lattice model.  On the other hand, in the case of $D=4$, the flat region 
disappears because the $\phi^4$ coupling is renormalized to 
zero.

Here we discuss a role of the effective potential in continuum quantum field 
theory.  After taking the continuum limit, 
it is believed that a minimum of the effective potential 
corresponds to the ground state (vacuum) in quantum field theory\cite{coleman}.  
If there are multiple points giving the minimum, we need to choose 
one point.  A standard way of doing this is to turn on a uniform 
external field as we demonstrated in sect.~3.   The vacuum selected 
in this way is homogeneous and corresponds to a point on the boundary of the 
flat region.   Since it  is the vacuum that is used in the previous 
literature\cite{cjp,kobayashikugo,fukuda,bm},  we can rederive the one-point function 
obtained in those works. However, physical observables concerned with higher-point correlation 
functions such as a mass matrix (or a susceptibility tensor) may be changed if we take 
the flat region into account.  We hope to report on this point more detail in a 
subsequent publication.  

In addition to reproducing the previous results, 
the effective potential obtained here 
suggests possibility of 
a non-uniform ground state such as 
a rotating configuration with a long-wave length in eq.(\ref{rotate}) 
or as kink ground states  
in the quantum ferromagnetic XXZ chain\cite{kn}. 
In order to study such a non-homogeneous vacuum, it is expected that 
a non-uniform external field will be needed, which will pick out a point 
inside the flat region. 

Related with this topic, 
we comment on the $1/N$ expansion in the presence of a non-uniform 
external field $\mbox{\boldmath $J$}_x$.  In this case, the action of 
the system changes to $S - \sqrt{N} \sum_{x} J^a_x \phi^a_x$.  The 
strategy adopted in sect.~2 leads us to the following saddle-point equation 
\beq
      \frac{1}{2} \sum_{x, y} 
              (\{ J_y^a + A^{-1}(\Omega \varphi^a - B^a) \} 
              \{ J_x^a + A^{-1}(\Omega \varphi^a - B^a) \} - A^{-1} ) G_{x z}G_{z y}
         + 
        \frac{1}{2} G_{z z} - 
        \frac{1}{g_0} \chi_z + \frac{m_0^2}{g_0}
      = 0,
\label{spj}
\eeq
where 
\beq
        G_{x y} \equiv \langle x | \frac{1}{-\Delta + \tilde \chi} | y \rangle,  \quad
        A \equiv \sum_{x,y} G_{x y}, \quad 
        B^a \equiv \sum_{x,y}  G_{x y} J^a_{y}.
\eeq
Although we need to solve this equation for carrying out the $1/N$ expansion, it is difficult 
because we cannot employ the ansatz of a uniform solution due to the first term  
\footnote{In ref.\cite{fukuda}, where a non-uniform external field is employed as a ``regularization'',
the first term is oversimplified and it misleads to a uniform solution. }.  In fact,  using the ansatz, 
we find that the constant solution $\chi^*$ must satisfy 
\beq
        \frac{1}{\Omega} \sum_{p \neq 0} \hat G_0(p) - \frac{2}{g} \chi^* + \frac{2 m_0^2}{g_0} 
        + (\frac{1}{\Omega} \sum_{p}  \hat{\cal  J}^a(p) \hat G_0(p) e^{ i p z})^2  = 0,   
\eeq
for an arbitrary site $z$.   Here $  \hat {\cal J}^a (p)$, $p \neq 0$ are identical with  Fourier components of 
$\mbox{\boldmath $J$}_x$
and 
\beq
        \hat  {\cal J}^a (0) \equiv \Omega \chi^* \varphi^a.   
\eeq
However it is obviously impossible if an oscillating mode of $\mbox{\boldmath $J$}_x$ is present.

Moreover,  a perturbative expansion in the external source 
fails even if the magnitude 
of it is infinitesimal because the expansion induces the infrared divergence 
in the broken phase.  

We feel that a constant mode of $\mbox{\boldmath $\phi_x$}$ is not appropriate 
for studying a non-uniform ground state. The order parameters cannot resolve 
whether an inner point of the flat region corresponds to an inhomogeneous states 
or a superposition of pure states selected by a uniform external source. 
Some oscillating modes will be chosen as order parameters.

\paragraph{Acknowledgments}
We are grateful to M.~Kato for useful discussions. 
We also would like to thank I. Ichinose for careful reading of the manuscript. 
One of us (H.~M.) is  grateful to  C.~Itoi,  H.~Tasaki and H.~Terao for fruitful discussions.  
He also thanks  T.~Koma and H.~Tasaki for explaining their works\cite{kt,kn}. 
Y.~S. wishes to thank R.~Fukuda, M.~Ikehara, M.~Ogata and F.~Sugino 
for enlightening discussions.

\end{document}